\documentclass[prb,twocolumn,floatfix,superscriptaddress,showpacs,nofootinbib]{revtex4}

\usepackage{amsfonts}
\usepackage{amssymb}
\usepackage{amsmath}
\usepackage{graphics}
\usepackage[pdftex]{graphicx}
\usepackage{color}

\newcommand{\half}{\frac{1}{2}}

\newcommand{\ket}[1]{{|{#1}\rangle}}

\begin{document}

%-----------------------------------------------------------------------------

\title{Experimentally Realizable C-NOT Gate in a Flux Qubit/Resonator System}
\author{Shiro Saito}
\email[email:]{s-saito@will.brl.ntt.co.jp}
\affiliation{NTT Basic Research Laboratories, NTT Corporation, 3-1 Morinosato-Wakamiya, Atsugi-shi, Kanagawa-ken 243-0198, Japan}
\author{Todd Tilma}
\email[email:]{ttilma@nii.ac.jp}
\affiliation{National Institute of Informatics, 2-1-2 Hitotsubashi, Chiyoda-ku, Tokyo-to 101-8430, Japan}
\author{Simon J. Devitt}
\affiliation{National Institute of Informatics, 2-1-2 Hitotsubashi, Chiyoda-ku, Tokyo-to 101-8430, Japan}
\author{Kae Nemoto}
\affiliation{National Institute of Informatics, 2-1-2 Hitotsubashi, Chiyoda-ku, Tokyo-to 101-8430, Japan}
\author{Koichi Semba}
\affiliation{NTT Basic Research Laboratories, NTT Corporation, 3-1 Morinosato-Wakamiya, Atsugi-shi, Kanagawa-ken 243-0198, Japan}

\date{\today}

\begin{abstract}
In this paper we present an experimentally realizable
microwave pulse sequence that effects a Controlled NOT (C-NOT) gate operation
on a Josephson junction-based flux-qubit/resonator system with high
fidelity in the end state. 
%In principle, our gate operation uses only the ground, and first excited state, of the resonator. 
We obtained a C-NOT gate process fidelity of $0.988$ ($0.980$) for a two (three) 
qubit/resonator system under ideal conditions, and a fidelity of $0.903$ for a two qubit/resonator
system under the best, currently achieved, experimental conditions. In both cases, we found that ``qubit leakage'' to 
higher levels of the resonator causes a majority of the loss of fidelity, and that such leakage becomes
more pronounced as decoherence effects increase.
 \end{abstract}


\maketitle

%-----------------------------------------------------------------------------
\section{Introduction\protect}\label{Introduction}

%It has been over a decade and a half since the first real quantum
%algorithm was proposed by David Deutsch and Richard
%Jozsa,~\cite{DeutschJoza1992} nearly two decades since quantum computing
%was seen to be a universal Turing process,~\cite{Deutsch1985,Deutsch1989} and over a quarter century since the eminent
%Richard Feynman initiated discussion on quantum
%computing.~\cite{Feynmann1982} However, the field of quantum computation is
%still under development, mostly  
%due to the difficulty of treating fragile quantum states coherently, 
%as well as a lack of a truly experimentally realizable scalable system.

The basic requirements for a successful quantum computer have been
expressed by David DiVincenzo nearly a decade ago.~\cite{DiVincenzo1995,DiVincenzo2000} 
These five criteria, listed below, have been widely accepted as being the best
road map for achieving realizable quantum computing by most research
programs throughout the world:
\begin{enumerate}
\item A scalable physical system of well-characterized qubits. \label{dc1}
\item The ability to initialize the state of the qubits to a simple fiducial 
state. \label{dc2} 
\item Long (relative) decoherence times, much longer than the gate-operation time. \label{dc3} 
\item A universal set of quantum gates. \label{dc4} 
\item A qubit-specific measurement capability. \label{dc5} 
\end{enumerate} 
At this time, there are various schemes being proposed to satisfy the
above criteria and realize a quantum computer.~\cite{Spiller2005} 
At the few-qubit level, these schemes include those based on trapped
ions,~\cite{CiracZoller1995} 
liner optics,~\cite{Knill2001,Kok2007} 
and nuclear spins in liquid-state molecules.~\cite{Cory1994,Gershenfeld1997}
For the long-term prospects of scalability though, those schemes that utilize Josephson
junction-based qubits~\cite{NEC1999} have significant advantages with current solid-state manufacturing technology.

Since the initial breakthrough in the coherent manipulation 
of a single Josephson junction-based charge qubit nearly a decade ago,~\cite{NEC1999} the
experimental focus has mostly shifted to the creation, control,
and subsequent manipulation of, 
multi-qubit entanglement in similar Josephson junction-based systems. 
For example, coherent oscillations between two qubits have been observed by using a 
fixed inter-qubit coupling.~\cite{NEC2003F,McDermott2005,Plantenberg2007}
However, the fixed nature of the qubit-qubit coupling used in these experiments makes it
difficult to scale up such circuits in the future. To overcome this
problem, a fast switchable 
coupling between two qubits has been proposed~\cite{Bertet2006,Niskanen2006} and also
demonstrated.~\cite{NEC2007}

Beyond direct qubit-qubit couplings, another solution is to make use of a quantum bus (qubus) as a coupler 
between qubits.~\cite{Blais2007,Nakano2007,BillKaequbus} Using the
``qubus'' concept, we can perform any two
qubit operation, between any two qubits that are coupled to the bus, 
without using multiple swap gates, which are necessary in other systems
using direct qubit-qubit coupling. In particular, harmonic oscillators formed by superconducting circuits
seem to be a good candidate for a ``qubus''-type coupler. Early
experiments with such couplers have shown 
coherent oscillations between a qubit and the oscillator, which was
made of lumped elements, namely capacitors and
inductors.~\cite{Chiorescu2004,Johansson2006} 
More recently, a distributed circuit, based 
on a coplanar waveguide 
resonator, attracted considerable attention as an oscillator because
of its high quality factor ($Q$-factor) and impedance matching to other
circuits.~\cite{Wallraff2004,Hofheinz2008} 
Furthermore, coherent quantum state transfer, between two Josephson junction-based qubits, via
such a waveguide resonator, has been demonstrated in both the phase~\cite{Sillanpaa2007} 
and charge regime.~\cite{Majer2007}
Lastly, more recently, two-qubit algorithms have been demonstrated in a two-transmon 
qubit/resonator system.~\cite{DiCarlo2009}

Because of the experimental viability of the ``qubus'' coupler
concept, as well as its obvious advantages in
scalability, we are using this paper to propose an experimentally realizable microwave pulse
sequence that will enact a controlled NOT (C-NOT) gate between two superconducting flux 
qubits, which are coupled via a harmonic oscillator bus. This pulse
sequence
flux qubit/resonator design is complementary to
other recently proposed systems,~\cite{Blais2007,Brito2008} 
but stands apart due to its
efficiency of operation and fidelity of its end state.

%-----------------------------------------------------------------------------
\section{Experimental System\protect}\label{Experiment}

One of the most promising solid state quantum computing elements is a
superconducting flux qubit that typically consists of 
three Josephson junctions in a loop: Two of equal size, one smaller by
a factor $\alpha \simeq 0.8$.~\cite{Mooij1999} 
The sizes of the junctions are chosen so that the geometric
self-inductance of the loop is not physically relevant. 
The two lowest energy states of the qubit at the flux degeneracy
point, usually denoted as $\ket{0}$ and $\ket{1}$, are superpositions
of macroscopically distinct clockwise and counter-clockwise persistent 
current states. Repeated experiments over the past few years have
shown that this type of flux qubit is a well-defined quantum system 
that can perform single-qubit
rotations~\cite{Chiorescu2003,Saito2006,Bertet2005,Yoshihara2006,Kakuyanagi2007} 
as well as achieve longer coherence times than other kinds of 
superconducting qubits.~\cite{Bertet2005} 
However, we have to operate the flux qubit at the degeneracy point in
order to enjoy the best coherence time because it is easily affected
by a flux fluctuation apart from the degeneracy point.~\cite{Bertet2005,Yoshihara2006,Kakuyanagi2007}

\begin{figure}[htbp]
\begin{center}
\includegraphics[width=1.0\linewidth]{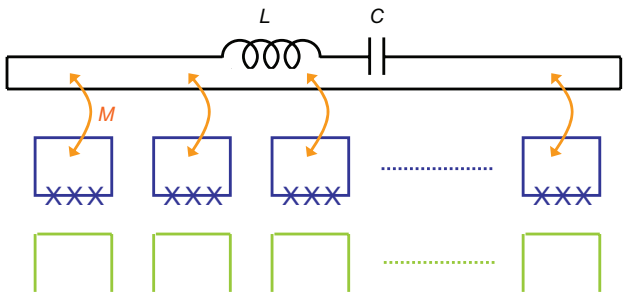}
\end{center} 
\caption{(Color online) Multiple flux qubits (blue rectangles with three crosses), 
each addressed by its own microwave line (green line), 
coupled to a resonator via a mutual inductance $M$. 
The resonator is schematically represented by the inductor $L$ and the capacitor $C$.}
\label{figure_circuit}
\end{figure}

With the above in mind, we are currently working with the following architecture
(see figure~\ref{figure_circuit}), 
built around the previously discussed three Josephson junction qubit design, 
which incorporates the fundamentals of the ``qubus''
concept.~\cite{BillKaequbus} 
Each qubit couples to a resonator through a mutual inductance $M$. 
Here the resonator is schematically represented by lumped elements,
but it can be a 
distributed circuit, for example, a coplanar strip waveguide. 
External magnetic flux through each qubit is a half flux quanta to set the qubits at the degeneracy point.

%-----------------------------------------------------------------------------
\section{C-NOT Pulse Sequence\protect}\label{PS}

In order to execute a C-NOT gate in our architecture (see figure~\ref{figure_circuit})
we have to entangle the flux qubits with the resonator. 
To do that,
we need to apply a sequence of DC-shift pulses, or apply a sequence of microwave pulses, 
through microwave lines to our qubits. 
In the case of the former pulses, we can adiabatically change the qubit frequency 
to fit the resonator frequency, but also non-adiabatically to create a coupling 
between the qubit and the resonator. As a result, we can ``turn on'' the coupling 
between the qubit and the resonator
non-adiabatically, making them into an entangled
state.~\cite{Johansson2006,Sillanpaa2007} 
However, these pulses cause DC-based excursions away from the flux
degeneracy point, and can reduce the dephasing times of the qubits 
drastically. 
The large bandwidth of the pulses can also reduce the overall gate
fidelity. Fortunately, a special flux qubit design may solve these problem by 
using elaborate DC pulses.~\cite{Brito2008} 

In the case of the latter
pulses, we can create entanglement between the qubits and the
resonator by using a known two-photon blue 
sideband (BSB) transition
at the qubit's degeneracy point.~\cite{Blais2007,Leek2008} 
These microwave pulses have a more
narrower bandwidth than the DC-shift pulses, thus allowing us to
obtaining a much higher fidelity gate, as well as minimizing pulse-induced dephasing.   
C-NOT gate operations based on such BSB transitions have been achieved 
with a high fidelity in ion trap experiments.~\cite{SchmidtKaler2003,Blatt2006} 
In these experiments, an elaborate controlled-$Z$ gate, 
using four BSB pulses, was used.~\cite{CiracZoller1995} 

\begin{figure}[htbp]
\begin{center}
\includegraphics[width=1.0\linewidth]{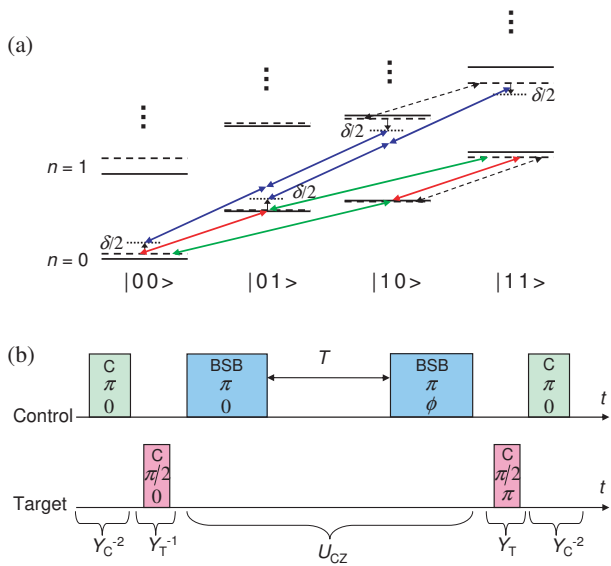}
\end{center} 
\caption{(Color online) 
(a) Energy levels of the two qubit/resonator system and transitions 
used in our pulse sequence. Solid (dashed) lines represent 
the energy levels with (without) the coupling between the qubits and 
the resonator. Additional shifts $\delta$ occur during the irradiation 
of the BSB pulse. Energy difference between the dashed arrows is utilized 
to perform the controlled-$Z$ gate between the resonator and the target qubit. 
(b) Basic operational sequence of ``target,'' ``control,'' and
BSB pulses on our two qubits, using values defined in
section~\ref{PSSimulation}, that enact a C-NOT
gate in our system. Each pulse is represented by the microwave frequency 
(Carrier frequency or BSB frequency), the rotation angle, and the phase of 
the microwave pulse. Note that the parameters $T$, the free evolution time, and
$\phi$, the microwave pulse phase, are variable: They are defined by
the experimental system to those values that optimize the operation of the gate.}
\label{pulse_sequence}
\end{figure}

It is an interesting idea to introduce such a BSB-based C-NOT gate operation
to our superconducting qubit/resonator system. However, it is hard  
to achieve a similar high fidelity because of the strong fixed coupling 
between the qubit and the resonator. This coupling 
leads to larger energy shifts at higher energy levels 
[see figure~\ref{pulse_sequence}(a)]. 
Indeed, the BSB-based C-NOT gate uses up to $n=2$ states, where $n$ is the photon 
number in the resonator. 
Therefore, we will utilize this coupling 
to realize a high-fidelity controlled-$Z$ gate instead 
by only using up to $n=1$ states. From this, we can build a 
pulse sequence that will do a C-NOT gate [figure~\ref{pulse_sequence}(b)].

As figure~\ref{pulse_sequence}(b) 
shows, the first BSB $\pi$ pulse transfers
information from the control qubit to the resonator, 
and the second one transfers the information back to the control
qubit. 
During a free evolution between the two pulses, we use only four energy 
levels, $\ket{10}$ and $\ket{11}$, for $n=0$ and $n=1$, where 
$\ket{CT}=\ket{C}\ket{T}$ and $\ket{C}$ ($\ket{T}$) 
represents the state of the control 
(target) qubit. The energy difference of the target 
qubit between $n=0$ and $n=1$ indicated by the two dashed arrows 
in figure~\ref{pulse_sequence}(a) enable us to realize 
a controlled-$Z$ gate between the resonator and the target qubit 
with a suitable temporal delay between the two pulses.
Similar gate operations have been employed in nuclear magnetic 
resonance quantum computation.~\cite{Vandersypen97}
These pulses, with proper phases, 
work as a controlled-$Z$ gate 
between the two qubits $(U_{\mathrm{CZ}})$.
We have to adjust the phase difference between the two pulses carefully 
to cancel the effect of the off-resonant ac-Stark shift $\delta$ 
in figure~\ref{pulse_sequence}(a).~\cite{Blais2007}
Finally, four single-qubit gates, 
and the $U_{\mathrm{CZ}}$ gate, form the C-NOT gate $(U_{\mathrm{C-NOT}})$ as follows:
\begin{equation}
U_{\mathrm{C-NOT}}=Y_{\mathrm{C}}^{-2} Y_{\mathrm{T}} U_{\mathrm{CZ}} Y_{\mathrm{T}}^{-1} Y_{\mathrm{C}}^{-2}=-
\left(
\begin{array}{cccc}
 0 & 1 & 0 & 0 \\
 1 & 0 & 0 & 0 \\
 0 & 0 & 1 & 0 \\
 0 & 0 & 0 & 1
\end{array}
\right)
\end{equation}
in the two-qubit basis $\{\ket{11},\ket{10},\ket{01},\ket{00}\}$. 
Here, $Y_{\mathrm{C}}$ and $Y_{\mathrm{T}}$ represent a $\frac{\pi}{2}$ rotation
of the control qubit, and that of the target one, respectively.

%-----------------------------------------------------------------------------
\section{Numerical Simulations\protect}\label{Simulation}

As has been noted recently,~\cite{Brito2008} in order to accomplish our
C-NOT, or any usable quantum computational gate, it will be necessary to 
manipulate multiple numbers of two-level quantum
states, which will be our physical qubits, at will with some type of
pulse sequence. For our system, as
discussed in section~\ref{PS}, and schematically represented in 
figure~\ref{figure_circuit}, this means that any single- or multi-qubit gate 
operation will be done using $\pi$
pulses, $\frac{\pi}{2}$ pulses, and so on, most likely delivered by an
on-chip microwave line, or lines. 
Our architecture not only achieves this, but also
allows for a Hamiltonian representation that easily allows us to 
simulate whatever necessary pulse sequences we
see fit to enact. In this section, our focus will be on simulating
those pulses that implement our high-fidelity C-NOT gate upon any two
qubits in the architecture.

To begin, when all the qubits in our system are at their respective
degeneracy points, the overall system is represented by the following Hamiltonian:
%\begin{equation}
\begin{eqnarray}
H &=& hf_{\textrm{res}}a^{\dagger}a
+\sum_{k}(\half h f_{k}\sigma_{z,k}+hg_{k}\sigma_{x,k}(a^{\dagger}+a) \\ \nonumber
&&+hA_{\textrm{MW},k}\sin(\omega_{k}t+\phi_{k})\sigma_{x,k}).
\label{hamiltonian}
\end{eqnarray}
%\end{equation}
Here, $h$ is the Plank constant, $f_{\textrm{res}}$ is the frequency of our
resonator, $\half h f_{k}\sigma_{z,k}$ represents our ``k-th''
qubit, $h g_{k}\sigma_{x,k}(a^{\dagger}+a)$ is the coupling term between the ``k-th''
qubit and the resonator,
and $hA_{\textrm{MW},k}\sin(\omega_{k}t+\phi_{k})\sigma_{x,k}$ describes the
microwave pulse for each qubit. 
Since we are only looking at C-NOT gates, we
will define one qubit as the ``target,'' and one as the 
``control'' qubit. Lastly, we should note that, in principle, our gate operation uses only the ground, and 
first excited state, of the resonator. 

Now, because we are wanting to analyze the time-evolution of multiple
qubits - in particular, whether or not our pulse sequence accurately
describes a C-NOT process on any two of them - we naturally will
want to work within the density matrix
formalism. In more detail, our simulated two-qubit gate, including
error sources, is properly described by a completely positive map
$\varepsilon$, 
the density matrix of 
the output state $\rho_\mathrm{{out}}$ of which can be written 
in the operator sum representation as~\cite{Chuang1997}
\begin{equation}
\rho_{\mathrm{out}}=\varepsilon(\rho_{\mathrm{in}})=\sum_{m,n} \tilde{E}_{m} \rho_{\mathrm{in}} \tilde{E}_{n}^{\dagger} \chi_{mn},
\label{chi}
\end{equation}
where $\tilde{E}_{m}$ are operators forming a basis in the space of
$4 \times 4$ matrices 
and $\rho_{\mathrm{in}}$ is the density matrix of the input state. 
This expression shows that $\varepsilon$ can be completely described
by a complex number matrix, $\chi$, once the set of operators
$\tilde{E}_{m}$ has been fixed. 

In this work, we have chosen these 16 operators to be
\begin{equation} 
\tilde{E}_{4i+j}=A_{i} \otimes A_{j} 
\end{equation} 
in the case of a two-qubit gate. Here, $A_{0}=I$, $A_{1}=\sigma_{x}$, 
$A_{2}=\sigma_{y}$, and $A_{3}=\sigma_{z}$. To determine $\chi$, which is a $16
\times 16$ matrix, 
we need to simulate our gate for 16 linearly independent input states
$\ket{\psi_{\mathrm{in}}}$. For these states we have chosen
\begin{align}
\ket{\psi_{\mathrm{in}}}&=\ket{\psi_{i}} \otimes \ket{\psi_{j}}, \\ \nonumber
\ket{\psi_{i}} &\in \{ \ket{0}, \ket{1}, (\ket{0}+\ket{1})/\sqrt{2},
(\ket{0}+i \ket{1})/\sqrt{2} \}
\end{align}
as the initial states. From this we can ``measure'' the gate fidelity using the process fidelity
\begin{equation}
F_{\mathrm{p}}=\mathrm{Tr}(\chi_{\mathrm{ideal}} \chi_{\mathrm{sim}}),
\label{fidelity}
\end{equation}
where $\chi_{\mathrm{ideal}}$ and $\chi_{\mathrm{sim}}$ represent the ideal matrix, and the one obtained from the simulation, respectively. 

%-----------------------------------------------------------------------------
\subsection{Simulation Parameters\protect}\label{PSSimulation}

Our two qubits (the ``target'' and ``control'') and resonator
operating frequencies, as well as couplings, 
were defined in the following way:
\begin{alignat}{2}
f_{1} &= f_{\textrm{Control}} = 6 \text{ GHz} &\qquad g_{1} &= 0.1 \text{ GHz} \\ \nonumber
f_{2} &= f_{\textrm{Target}} = 5 \text{ GHz} &\qquad g_{2} &= 0.1 \text{ GHz} \\ \nonumber
f_{\textrm{res}} &= 10 \text{ GHz}.
\end{alignat}
We then generated numerical simulations of
equation~\eqref{hamiltonian} using the following values for our
target and control pulses: 
\begin{alignat}{2}
&\textsc{Target Pulse} &\qquad &\textsc{Control Pulse} \\ \nonumber
& A_{\textrm{MW},2 } = 0.1 \text{ GHz} &\qquad &A_{\textrm{MW},1} = 0.1 \text{ GHz} \\ \nonumber
&\omega_{2}/2\pi = 4.99875 \text{ GHz} &\qquad &\omega_{1}/2\pi = 5.9981 \text{ GHz} \\
&\phi = 0 \text{ or } \pi &\qquad &\phi = 0, \nonumber
\end{alignat}
as well as our blue-side-band (BSB) pulses:
\begin{alignat}{1}
&\textsc{Control BSB} \\ \nonumber
&A_{\textrm{MW},1} = 2 \text{ GHz} \\ \nonumber
&\omega_{1}/2\pi = 7.5601 \text{ GHz} \\
&\phi = 0 \text{ or } 0.34\pi. \nonumber
\end{alignat}
Here the rise/fall time of each pulse was set at $0.8 \text{ ns}$ and the duration of the
carrier $\pi$($\frac{\pi}{2}$)-pulse was set at $5(2.5) \text{ ns}$. 
As we mentioned in section~\ref{PS}, we optimized the duration of 
the BSB pulse, the free evolution time $T$, and the phase of the second
BSB pulse $\phi$ in order to achieve the best gate fidelity; obtaining
$16.295 \text{ ns}$, 
$162.865 \text{ ns}$ and $0.34 \pi$, respectively. The total length of
the pulse sequence is about $200 \text{ ns}$, which is much shorter than the
reported coherence time of a flux qubit.~\cite{Bertet2005} Lastly, we assumed 
that the first five resonator levels were
accessible, thus $n=0$ to $n=4$, where $n$ is the photon number in the resonator. 

%-----------------------------------------------------------------------------
\section{Observation and Analysis\protect}\label{PSAnalysis}

Multiple simulations, using the previously defined parameters with various pulse shapes
and operation times, have yielded an experimentally 
realizable matrix of target, control and BSB pulse sequences,
represented in figure~\ref{pulse_sequence}(b), that will initiate a C-NOT gate upon our 
two qubit/resonator system with high process fidelity,
$F_{\mathrm{p}}=0.988$, 
and within the coherence time of known flux
qubit systems. The computed gate is as follows (to four significant figures):
\begin{equation}
U^{\prime}_{\mathrm{C-NOT}}=\left(
\begin{array}{cccc}
 0.0001 & 0.9863 & 0.0013 & 0.0022 \\
 0.9863 & 0.0001 & 0.0020 & 0.0013 \\
 0.0021 & 0.0014 & 0.9886 & 0.0034 \\
 0.0013 & 0.0023 & 0.0035 & 0.9880
\end{array}
\right).
\end{equation}
%
%\begin{figure}[htbp]
%\begin{center}
%\includegraphics[width=1.0\linewidth]{figure3.eps}
%\end{center} 
%\caption{Exact ($U_{\mathrm{C-NOT}}$), left, and realized ($U^{\prime}_{\mathrm{C-NOT}}$), right, two-qubit C-NOT gate via
%the representative pulse sequence shown in figure~\ref{pulse_sequence}. The realized
%C-NOT gate has an process fidelity greater than 98\% without decoherence. The basis
%corresponds to the situation where there was no photon in the resonator ($n=0$). The color
%scheme used is a nonlinear gradient.}
%\label{figure_cnot}
%\end{figure}

\begin{figure}[htbp]
\begin{center}
\includegraphics[width=0.8\linewidth]{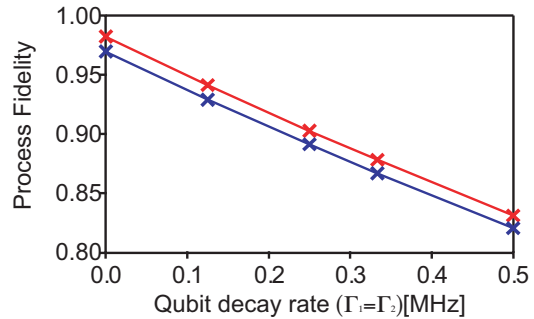}
\end{center} 
\caption{(Color online) Process fidelity as a function of qubit decay rate. 
The upper (lower) curve represents the fidelity when the quality factor 
of the resonator is $10^{6}$ ($3\times10^{5}$).}
\label{decoherence}
\end{figure}

\begin{figure}[htbp]
\begin{center}
\includegraphics[width=1.0\linewidth]{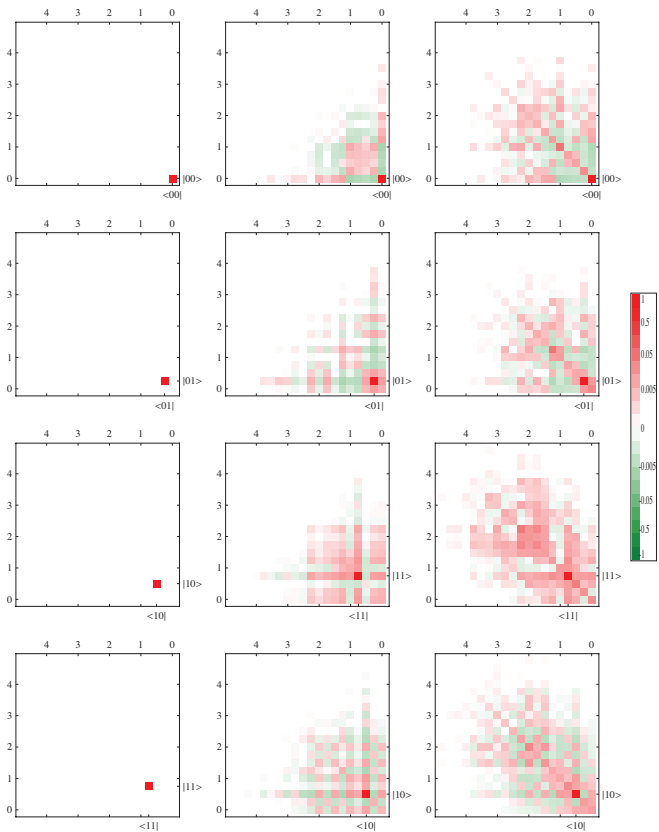}
\end{center} 
\caption{From top: Initial (left) and final (middle and right) density matrices (real components only) showing the $\ket{00} \mapsto \ket{00}$, $\ket{01} \mapsto \ket{01}$,
$\ket{10} \mapsto \ket{11}$, and $\ket{11} \mapsto \ket{10}$ operations, and 
the corresponding qubit leakage into higher resonator modes ($n>0$), due to the pulse
sequence values defined in section~\ref{PSSimulation}, without and with decoherence 
($Q~=~10^{6}$, $\Gamma_{1} = \Gamma_{2} = 0.25 \text{ MHz}$). The color
scheme used is a nonlinear gradient: Zero values are white, with negative values greenish and positive values reddish.}
\label{figure_combined}
\end{figure}

We also evaluated the effects of decoherence on our C-NOT gate 
by introducing a linear loss to the resonator (quality factor $Q$), 
as well as relaxation rates $\Gamma_{1}$ and dephasing rates 
$\Gamma_{2}$ to the qubits, 
via a master equation of the Lindblad form (see figures~\ref{decoherence} and~\ref{figure_combined}). 
Here we assumed that $\Gamma_{1}$ and $\Gamma_{2}$ for both qubits 
were equal. We obtained a process fidelity of $0.903$ in the best 
conditions which have been achieved experimentally, for example, 
$Q=10^{6}$, $\Gamma_{1} = \Gamma_{2} = 0.25 \text{ MHz}$. 
Although the fidelity without the decoherence is not unity, 
it is much better than that with the decoherence, hence our pulse 
sequence could be useful enough to demonstrate the C-NOT gate experimentally.
%From now on, we will not consider decoherence when discussing the fidelity. 

Now, in the case of flux qubits, it is difficult to fabricate a qubit 
which has an exact designed gap frequency at the degeneracy point, 
unless using the qubit with a controllable third junction.~\cite{Paauw2009} 
Hence, we also simulated the case in which the two qubit frequencies became 
closer to each other. When $f_{\textrm{res}} = 10 \text{ GHz}$ and 
$f_{\textrm{Control}} = 6 \text{ GHz}$ and $f_{\textrm{Target}} = 5.5 \text{ GHz}$, 
we obtained a process fidelity of $0.986$ without decoherence. 
Decreasing the difference between two qubit frequencies did not 
significantly affect the fidelity.

This gate also works with any two qubits in a system with many qubits, 
since we can decouple any unused qubits from the resonator 
via very quick ($\approx 5 \text{ ns}$) $\pi$ pulses applied at the
center of the two BSB pulses. For example, we obtained an process
fidelity of $F_{\mathrm{p}}=0.980$ 
in a three qubit/resonator system 
where the frequency of the third unused qubit was $7 \text{ GHz}$,
its coupling to the resonator 
was $0.1 \text{ GHz}$, and the microwave frequency of the decoupling 
pulse was $6.9973 \text{ GHz}$. The other system conditions were
similar to that use in the two-qubit simulation. 

Unfortunately, our gate fidelity, while still better than that currently achieved by other 
experimental systems (see for example Riebe \emph{et.\ al. }and their
table II),~\cite{Blatt2006} is 
still not unity. The lack of unit fidelity in our C-NOT gate is due to ``qubit
leakage''; the unintended, and persistent, excitation of higher-order
resonator levels. As Fazio
\emph{et.\ al. }noted in 2000,~\cite{Fazio2000}
qubit leakage issues are due to the fact that the qubit Hilbert space is actually
a subspace of a larger Hilbert space; in our case, one that includes
accessible higher-order resonator levels.
Although their work focused on charge qubits, their basic conclusions hold for flux qubits as well, as
our simulations have shown (see
%figures~\ref{figure1} through~\ref{figure4}).
figure~\ref{figure_combined}).

In more detail, coherent leakage into higher order modes of the
harmonic oscillator bus seems to be responsible for the missing process fidelity of the system.  Not only is this large an 
error rate unacceptable from the standpoint of large scale, fault-tolerant quantum computation, but the 
error channel itself is problematic to correct under current quantum error correction (QEC) schemes.

One of the fundamental assumptions of 
QEC is that errors affecting qubits are localized to the two-levels associated with 
the qubit encoding. Once errors begin to violate this assumption, either through state leakage or actual 
qubit loss, additional mechanisms must be employed, at the encoding level, to allow for correction.
Many different techniques have been developed to compensate for this type of error channel, including 
leakage protection through non-demolition detection and state
re-injection,~\cite{det1,det2,det3,det4} correction by teleporting 
``leaked'' qubits back into the 2-level subspace,~\cite{tele1} and specialized leakage reduction units (LRU's) which 
suppress leakage through a pulse sequences in a similar manner to bang-bang protection for 
qubit dephasing.~\cite{Lidar1,Lidar2} Unfortunately, these
techniques are quite cumbersome and involve specialized protocols that go beyond the 
standard operations to realize QEC. 

Ideally, for large scale qubit applications, we do not want to 
correct for qubit leakage, instead we wish to suppress it to a sufficient level where it can be ignored. It should be noted that 
this requirement is not as trivial as it might appear. For large scale applications, fault-tolerant quantum computation 
requires extremely low error rates. This ``target'' error rate for fault-tolerant QIP, often referred to as the ``fault-tolerant threshold'' 
is highly dependent on the underlying physical architecture and can be anywhere between $10^{-3}$ and 
$10^{-7}$.~\cite{aliferis1,ash1} The important point to realize is that if we wish to ignore coherent leakage errors, and 
avoid implementing specialized protocols, errors caused by qubit leakage will need to be several 
orders of magnitude below any other correctable source of qubit error. Therefore, the pulse design of this 
bus mediated qubit/qubit coupling will be required to exhibit effectively zero leakage, even before open system quantum 
effects are even considered, if it is to be used to do realizable QIP. This means that a new pulse sequence will need
to be created.

%-----------------------------------------------------------------------------
\section{Follow-On Pulse Sequence Possibilities\protect}\label{GS}

From our work, it is our belief that the seen qubit leakage, even taking our simple decoherence model into account, 
and the corresponding non-unity fidelity is an unavoidable consequence 
of the programmed waveforms, and expected physical implementation, of our microwave pulses. Experimentally,
in order to avoid qubit state transitions, the rise and fall times of our pulses have to be long enough
to fulfill the adiabatic condition, but also short enough to avoid
unwanted relaxation processes. 
Temporal lengths of the various target, control and BSB pulses used also 
depend on the microwave pulse amplitude, and are experimentally determined from the system's 
driven Rabi oscillations.
Thus, while these numerical simulations demonstrate the principal application of 
a bus-mediated controlled qubit/qubit interaction, the implementation of the proposed pulse sequence 
requires further improvement of the leakage to build up not only a several qubit 
system, from the standpoint of a quantum testbed, but also a really-scalable QIP system.

In more detail, analysis of our control, target and BSB pulse matrices indicates that our qubit leakage is
initiated by our first BSB pulse on the control qubit. 
Since the second BSB pulse on the same qubit is only different by a phase, 
we only require a modified version of the BSB pulse matrix to avoid leaking 
population into higher order modes of the harmonic oscillator. 
Also, as the simple ``top hat'' pulse design we have examined is already very well confined, it is expected that 
this pulse represents a good first approximation to a zero leakage modification, and due to the structure of this 
sequence, we already have a good idea which section of the pulse is
``leaky.'' Thus we feel that the only way to truly generate a microwave pulse-induced
gate, with fidelity experimentally equivalent to unity, for this system will be to use
technologies based on optimal quantum control.~\cite{Judson1,Krotov1,Khaneja,Schulte1,Schirmer1} 

Optimal quantum control is a vast area of theoretical and numerical
techniques that were 
developed largely in the field of nuclear magnetic resonance (NMR) and quantum chemistry, in order to 
find complicated control fields for NMR manipulations, or to increase chemical yields in chemical reactions 
involving photonic reagents.  
This research has more recently been applied to the quantum
information
field;~\cite{Schulte1,Schulte2,Tesch,Rebentrost1,Nebendahl1} utilized 
to construct high-fidelity quantum gates from a set of classical controls, which may be 
pulsed in counter-intuitive ways, that would be difficult to find with
traditional analytical techniques.  

There are many different ways in which to formulate the optimal control problem, in terms of the target quantum operation, and 
in terms of numerical search methods. 
In a very broad way, the general idea of optimal control methods is to
iteratively search for a target set of control field parameters that drive 
a system Hamiltonian in a controlled manner. Depending on the system considered, the target parameter could be quantum 
process fidelity, or operational time of the evolution, or the value of a specific quantum 
observable. 

One such example of an optimal control method is the
standard numerical algorithm known as gradient ascent pulse
engineering (GRAPE);~\cite{Khaneja} 
optimal quantum gates (in terms of process fidelity) are constructed by iteratively varying 
control parameters, and analytically calculating the gradient changes,
in order to find the minimum fidelity points in 
the underlying unitary space. These techniques have been applied to ion traps,~\cite{Rebentrost1} vibrational modes in
molecules,~\cite{Tesch} as well as in the superconducting regime~\cite{WilhelmGlaser2007,Rebentrost1}
with significant success.

Our future work will involve utilizing these numerical techniques to find smooth control field parameters 
that modify the above simulated BSB pulse to one that eliminates leakage to higher order field modes.  
Hopefully, once we eliminate the final issue of qubit leakage, we will utilize these numerical algorithms in 
full open system calculations to develop high fidelity control pulses for superconducting systems 
undergoing decoherence.

%-----------------------------------------------------------------------------
\section{Conclusions\protect}\label{Conclusion}

Our simulations have confirmed a usable control, target, and BSB pulse
matrix, as well as the required operational pulse rise/fall time
envelope published previously,~\cite{condmat07104455}
for an experimentally successful microwave pulse-induced C-NOT gate.
However they have also shown that this set of pulses, in particular
the BSB set,
incites unwanted qubit leakage, at a minimum, into the first and second
excited states of the coupling resonator, which makes them troublesome
for usage in a more QIP-focused architecture. 

Our proposed solution to this problem, replacing the BSB pulse
sequence with a sequence based on optimal quantum control concepts, seems to have the
potential to reduce the observed qubit leakage. Yet it is still uncertain whether or not the required
pulse sequences needed for these codes can be realized
experimentally. The main reason is that our current architecture has a 1:1
resonance between qubit number and microwave lines. This was done in order to avoid
cross-talk between qubits. More complicated
pulse sequences on this architecture, such as those needed for GRAPE codes, run
the risk of initiating cross-talk between qubits and other microwave lines, or exciting on-chip fluctuators close to, or within, each
qubit's junction. 

Currently, we are planning to explore
these issue in more detail by investigate other GRAPE-inspired multi-qubit/resonator gate sequences, as well as modifications
to the architecture.
Future work will also look into more accurately modeling the dynamical impact of the various low- and high-
frequency noise sources inherent in our system,~\cite{condmat07104455}
using a more generalized open quantum
system treatment.~\cite{Cesar2008}  

%-----------------------------------------------------------------------------
\section{Acknowledgments\protect}\label{Acknowledgments}

We would like to thank W. J. Munro, H. Nakano, T. P. Spiller, and J. E. Mooij for numerous, useful
discussions and The Center for Complex Quantum Systems at The University of Texas at Austin
for their continued support and encouragement.
This work was supported in part by Grant-in-Aid for Scientific Research
of Specially Promoted Research \#18001002 by MEXT, Grant-in-Aid for
Scientific Research (A) \#18201018 by JSPS.

%-----------------------------------------------------------------------------
%
% ***** Bibliography *****
%

%-----------------------------------------------------------------------------

%
% ***** end of file base.tex *****

\end{document}